%% file: 0-main.tex
\documentclass{bioinfo}
\copyrightyear{2021} \pubyear{2021}

\access{Advance Access Publication Date: Day Month Year}
\appnotes{Manuscript Category}

\usepackage{algorithm}
\usepackage{graphicx}
\usepackage{amsmath}
\usepackage{subfigure}
\usepackage{soul,color}
\usepackage{algorithm,algpseudocode}
\usepackage{url}
\usepackage{breqn}
\usepackage{caption}
\usepackage{amsthm}
\theoremstyle{definition}
\newtheorem{theorem}{Theorem}[section]
\newtheorem{lemma}{Lemma}[section]

\newenvironment{prooF}{\paragraph{Proof:}}{\hfill$\square$}
\usepackage{titlesec}
\usepackage{xcolor}

\begin{document}

\bibliographystyle{bioinformatics}
\firstpage{1}
\subtitle{Genome Analysis}

\title[GenShare]{GenShare: Sharing Accurate Differentially-Private Statistics for Genomic Datasets with Dependent Tuples}
\author[Almadhoun Alserr \textit{et~al}.]{Nour Almadhoun Alserr\,$^{\text{\sfb 1,3}}$, Ozgur Ulusoy\,$^{\text{\sfb 3}}$, Erman Ayday\,$^{\text{\sfb 2,}*}$, and Onur Mutlu\,$^{\text{\sfb 1,3}*}$}
\address{$^{\text{\sf 1}}$Department of Information Technology and Electrical Engineering, ETH Zurich, Zurich 8006, Switzerland\,  $^{\text{\sf 2}}$Department of Electrical Engineering and Computer Science, Case Western Reserve University, Cleveland, OH 44106, USA\, $^{\text{\sf 3}}$Computer Engineering Department, Bilkent University, Ankara 06800, Turkey}
\corresp{$^\ast$To whom correspondence should be addressed.}

\history{Received on XXXXX; revised on XXXXX; accepted on XXXXX}

\editor{Associate Editor: XXXXXXX}

\abstract{\textbf{Motivation:} Cutting the cost of DNA sequencing technology led to a quantum leap in the availability of genomic data. While sharing genomic data across researchers is an essential driver of advances in health and biomedical research, the sharing process is often infeasible due to data privacy concerns. Differential privacy is one of the rigorous mechanisms utilized to facilitate the sharing of aggregate statistics from genomic datasets without disclosing any private individual-level data. However, differential privacy can still divulge sensitive information about the dataset participants due to the correlation between dataset tuples.\\
\textbf{Results:} Here, we propose GenShare model built upon Laplace-perturbation-mechanism-based DP to introduce a privacy-preserving query-answering sharing model for statistical genomic datasets that include dependency due to the inherent correlations between genomes of individuals (i.e., family ties). We demonstrate our privacy improvement over the state-of-the-art approaches for a range of practical queries including cohort discovery, minor allele frequency, and ${\chi}^2$ association tests. With a fine-grained analysis of sensitivity in the Laplace perturbation mechanism and considering joint distributions, GenShare results near-achieve the formal privacy guarantees permitted by the theory of differential privacy as the queries that computed over independent tuples (only up to 6\% differences). GenShare ensures that query results are as accurate as theoretically guaranteed by differential privacy. For empowering the advances in different scientific and medical research areas, GenShare presents a path toward an interactive genomic data sharing system when the datasets include participants with familial relationships.\\
\textbf{Contact:} \href{exa208@case.edu}{ exa208@case.edu}, \href{exa208@case.edu}{omutlu@ethz.ch}\\
\textbf{Supplementary information:} Supplementary data are available at \textit{Bioinformatics} online.}

\maketitle

\input{1-Introduction}

\input{2-RelatedWork}
\input{3-Method}
\input{4-Results}
\input{5-Conclusion}

\end{document}

%% file: 1-Introduction.tex
\section{Introduction}
\bibliographystyle{bioinformatics}
The fast-paced high throughput sequencing technologies result in generating a tsunami of large-scale datasets and biobanks. The number of sequenced human genomes has been increasing at an exponential rate, and now we are at about 2.5 million sequenced genomes around the world. This is projected to reach 105 million genomes in 2025~\cite{stephens2015big}, especially after the COVID-19 pandemic, where many countries have decided to study genomic data at a population scale. %which increases the probability of having families in the genomic datasets.
These rich troves of data are becoming the keystone for empowering medical science advances. Researchers need large amounts of \emph{genomic datasets} that they can leverage to gain a better understanding of 1) the genetic basis of the human genome and identify associations between phenotypes and specific parts of DNA, and 2) disease diagnosis and treatment (e.g., personalized medicine~\cite{farnaes2018rapid}). %divulging
However, since the \emph{human genome} is the utmost personal identifier, it is normally discouraged to share genomic data due to the \emph{privacy} concerns and the possible legal, ethical, and financial consequences, as well as the data protection guidelines in many countries. Hence, sharing genomic data while preserving the privacy of the individuals has been challenging for many different fields (e.g., medicine, science, bioinformatics)~\cite{bonomi2020privacy}. The challenge worsens when sharing large datasets or their statistics as they are usually vulnerable to privacy leaks due to the inherent correlations between genomes of participating family members~\cite{almadhoun2020differential,almadhoun2020inference}. 
%Hence, due to data privacy concerns, sharing these data across repositories is often infeasible. 
For the hope of sharing genomic datasets and aiming at gaining more accurate and refined biomedical insights, researchers have proposed applying \emph{differential privacy} (DP) concept~\cite{dwork2008differential} as a protective measure against several inference attacks over genomic dataset (e.g., Homer attack~\cite{homer2008resolving}). %DP, proposed over a decade ago, has become a primary standard for privacy.
Informally, a (\emph{randomized}) algorithm $A$ is differentially private if its output distribution is approximately the same when executed on two inputs (e.g., datasets $D$ and $D'$) that differ by the presence of a single individual’s data (i.e., \emph{neighboring datasets}). This condition prevents an adversary with access to the algorithm output from learning anything substantial about any one individual since the probability of observing a certain outcome for the neighboring datasets does not differ by more than a multiplicative factor of $exp$ $\epsilon$. $\epsilon$ is referred to as the \emph{privacy budget}, where smaller values of $\epsilon$ give stronger guarantees of privacy. DP methods are widely-used for privately sharing the summary statistics after adding an adequate noise. One of the DP common approaches is to add Laplace noise (i.e., \emph{Laplace perturbation mechanism} (LPM))~\cite{nissim2007smooth} based on the \emph{global sensitivity} (GS) of the statistics query (i.e., the maximum difference between the query results $A(D)$ and $A(D')$ is at most $GS(A)$).
\cite{uhlerop2013privacy,yu2014scalable,johnson2013privacy} developed differentially-private algorithms that release different queries in a privacy-preserving way from statistical genomic studies, such as \emph{genome wide association studies} (GWAS). These queries include but are not limited to 1) \emph{count} or \emph{cohort discovery}: to query how many participants in the dataset satisfy given criteria, 2) ${\chi}^2$ association tests: compute ${\chi}^2$ statistics for a point mutation (\emph{single nucleotide polymorphism} SNP), or 3) \emph{minor allele frequency} (MAF): to compute the frequency of which the rare nucleotide occurs at a particular SNP.
%Non-negative real number e is referred to as the privacy parameter, where smaller values of e give stronger guarantees of privacy

%To address these attacks,current systems and studies have attempted to apply differential privacy mechanism to perturb the query results with a small amount of noise in order to reduce the individuals privacy concerns.

Despite the rigorous mathematical foundation of DP~\cite{cho2020privacy,he2018achieving,raisaro2018m} and the fact that only aggregate-level information is shared, DP mechanisms can still leak sensitive information about the participated individuals if the dataset includes \emph{dependent tuples} (i.e., family members). It is a common situation for genomic datasets to have dependency between their tuples (or records) due to the inherent \emph{correlations} between genomes of individuals that have family ties. In our previous work~\cite{almadhoun2020differential,almadhoun2020inference}, we demonstrate the feasibility of attribute and membership inference attacks on differentially private query results by exploiting the dependence between tuples. Our evaluation over real-world statistical genomic datasets shows how kinship relations between individuals participating in a genomic dataset cause a significant \emph{reduction} in the privacy guarantees of traditional DP-based mechanisms. Current studies have attempted to propose general mechanisms to tackle this problem, such as Pufferfish~\cite{kifer2011no}, and its extensions~\cite{he2014blowfish,chen2014correlated,liu2016dependence,zhao2017dependent}. However, these efforts fail to capture the \emph{statistical relationships} between dependent tuples in genomic datasets, and hence resulting in sub-optimal solutions limiting their effectiveness in practice. They either lack the \emph{privacy} (degrade rigorous guarantees of privacy) or the \emph{utility} (introduce an excessive amount of noise), as we show in our evaluation in Section~\ref{EvaluationSec}. Therefore, there is a critical need for a fine-grained \emph{analysis} of LPM sensitivity considering different queries over genomic datasets to fill an unmet need for privacy-preserving genomic data sharing when the dataset includes dependent tuples. This will encourage both the healthcare stakeholders and data donors (including families) to widely share and use such valuable data resources. 

Our $\textbf{goal}$ in this paper is to enable privacy-preserving sharing of summary statistics from genomic data with dependent tuples by achieving the privacy and utility (encompassing \emph{accuracy}) guarantees of the standard DP assuming all the participants of the dataset are \emph{independent} (i.e., independent tuples). To achieve this goal, we aim at preserving the privacy of the genomic data donors by analyzing and perturbing the query results using a controlled noise in order to minimize the probability of leaking undesired information. We propose $\textbf{GenShare}$ model that provides rigorous theoretical guarantees of DP formulation in terms of privacy and utility. The $\textbf{key idea}$ of GenShare is to 1) theoretically analyze statistical relationships between the tuples in the genomic datasets to infer both pairwise correlations and complex joint correlations between multiple participants, 2) compute the dependence sensitivity ($\sigma$) sensitivity, how much each query can reveal out of such statistical relationships, and 3) take effective DP protective measures based on each query sensitivity. Focusing on three types of real-world queries: (1) count or cohort discovery, (2) MAF, and (3) ${\chi}^2$ tests, we empirically demonstrate the privacy and utility improvements of our proposed DP-based mechanism for each query type. We provide a use case on how our GenShare could be used to enable data sharing with privacy.
%In GenShare we leverage the LPM for achieving differentially private result for each query type. 
Our key theoretical advances show that an LPM based approach, combined with a fine-grained computation for the sensitivity performed by the \emph{data owner} (i.e., entity which collects/generates the genomic dataset), provably achieves the expected data utility of the shared query results, while maintaining the privacy guarantees of DP that can be obtained when the query is computed over independent tuples. This paper makes the following  $\textbf{contributions}$:
\begin{itemize}
    \item Introducing a query-answering sharing model “GenShare" for genomic datasets with formal privacy guarantees, while ensuring that the query results are as accurate as theoretically guaranteed by DP.
    \item Providing an effective LPM-based analysis based on the dependent and independent tuples included in the query computations, which is more accurate and robust than most similar existing approaches.
    \item Following the real-world workflows in recent studies for different queries, we show the robustness of GenShare using a range of queries such as cohort discovery, MAF, and ${\chi}^2$ over real-world statistical genomic datasets.
    \item Achieving almost the same privacy guarantees (in terms of estimation error, which is commonly used to quantify genomic privacy) as the query that is computed over independent tuples.
\end{itemize}

To our knowledge, GenShare is the first model that 
%shows how one can leverage the theory of DP and analyze the LPM sensitivity to protect the privacy of individuals in the genomic dataset considering \emph{joint} correlations,
dynamically and effectively tailors the DP protective measures based on each query sensitivity to protect the privacy of individuals who have simple/complex correlations participating in the genomic dataset, while simultaneously maximizing the benefits of data sharing for science. The rest of this paper is organized as follows: Section 2 presents related prior work on DP mechanisms under dependent tuples. Section 3 explains our proposed privacy model “GenShare", followed by Section 4 where we evaluate our proposed GenShare model and compare it to the state-of-art mechanisms. Section 5 presents the conclusion and highlights future research directions that are pointed by this paper.

%% file: 2-RelatedWork.tex
\section{Related Work} \label{RelatedWorkSec}

%Considering two types of correlations: 1- the dependency between different tuples (e.g. NDSS (2016) for genomic statistical datasets including family members) 2- correlated time-series (streams) entries (e.g. pufferfish (2017) and pegasus (2017) for physical activities of a single entry across time). \newline

Several studies have questioned whether DP is valid for correlated data. \cite{kifer2011no} was the first to raise the issue of privacy degradation when DP is applied over a dataset with correlated tuples. To this end, existing solutions that try to handle the correlation between tuples in the datasets can be categorized into two types, by considering: 1) the dependency between different tuples (i.e., individual-individual correlations), and ii) the dependency among single individual’s data at different time-series (streams) entries (i.e., temporal correlations).

First, to handle the individual-individual correlations (or vertical correlations) between tuples, Group DP~\cite{dwork2014algorithmic} is one of the first studies, which proposes adding noise proportional to the size of the largest correlated tuples in the dataset. Their method adds a tremendous amount of noise (i.e., $O(b)$ noise to a dataset with $b$ dependent tuples), thus destroying the data utility. As a generalization of DP, \cite{kifer2012rigorous} proposes another general and customizable method called Pufferfish to handle the dependent tuples by adjusting the Laplace scale, however, the main challenge of Pufferfish is the lack of suitable mechanisms to achieve the expected privacy guarantees. Following this general approach of Pufferfish, the baseline approach proposed by~\cite{chen2014correlated} tries to handle the correlation by multiplying the original sensitivity of the query with the number of correlated records $b$ (i.e., query sensitivity = $b$ $\times$ query original sensitivity). Bayesian DP~\cite{yang2015bayesian} uses a modification of Pufferfish, but it only focuses on modeling the tuples correlation by the Gaussian Markov Random Fields. All the following studies such as~\cite{liu2016dependence,zhao2017dependent,almadhoun2020differential} are trying to adjust the sensitivity by introducing dependence coefficients according to the number of correlated data, considering the pairwise correlation between dataset tuples as in~\cite{liu2016dependence} or using heuristic analysis (empirically-computed query sensitivity) as in~\cite{almadhoun2020differential}.

Following the second setting to handle the temporal correlations,~\cite{song2017pufferfish,chen2017pegasus} propose sharing statistics and counts of a data stream considering horizontal correlations. In~\cite{song2017pufferfish}, they propose two algorithms for the Wasserstein mechanism and the Markov Quilt mechanism when the correlations can be modeled by Bayesian Network. \cite{cao2017quantifying} also considers the temporal correlation which can be modeled by a Markov Chain.

In Section~\ref{ExperimentalResultsSec}, we compare our model (in terms of privacy) with the existing similar approaches from the two aforementioned categories~\cite{liu2016dependence,almadhoun2020differential,song2017pufferfish,dwork2014algorithmic}. Since Hidden Markov would not work to model %kinship relations in a 
statistical genomic dataset, we are not comparing our model with the mechanisms proposing hidden Markov-based models~\cite{yang2015bayesian,song2017pufferfish,chen2017pegasus}.
% Since Hidden Markov Model would not work to model kinship relationships in a data stream, we are comparing our model (using statistical genomic datasets) with the Wasserstein mechanism.  

%- limitations of current works: 
%As we demonstrate in our work, existing methods perturb the results more or less than is necessary to achieve a desired level of privacy.  We are trying to show that privacy obtained with this noise becomes even stronger than the privacy obtained by traditional DP over independent tuples. We quantify this privacy using the correctness or estimation error metric. In other words, such standard approaches provide more privacy than required, and hence they reduce utility. On the other hand, we model correlations in such a way that the privacy we achieve becomes very similar to the one achieved with independent tuples. We also have to show that they never compare with the privacy achieved with independent tuples \newline

%% file: 3-Method.tex
\begin{figure*}[b]
\center
\includegraphics[width=0.8\textwidth]{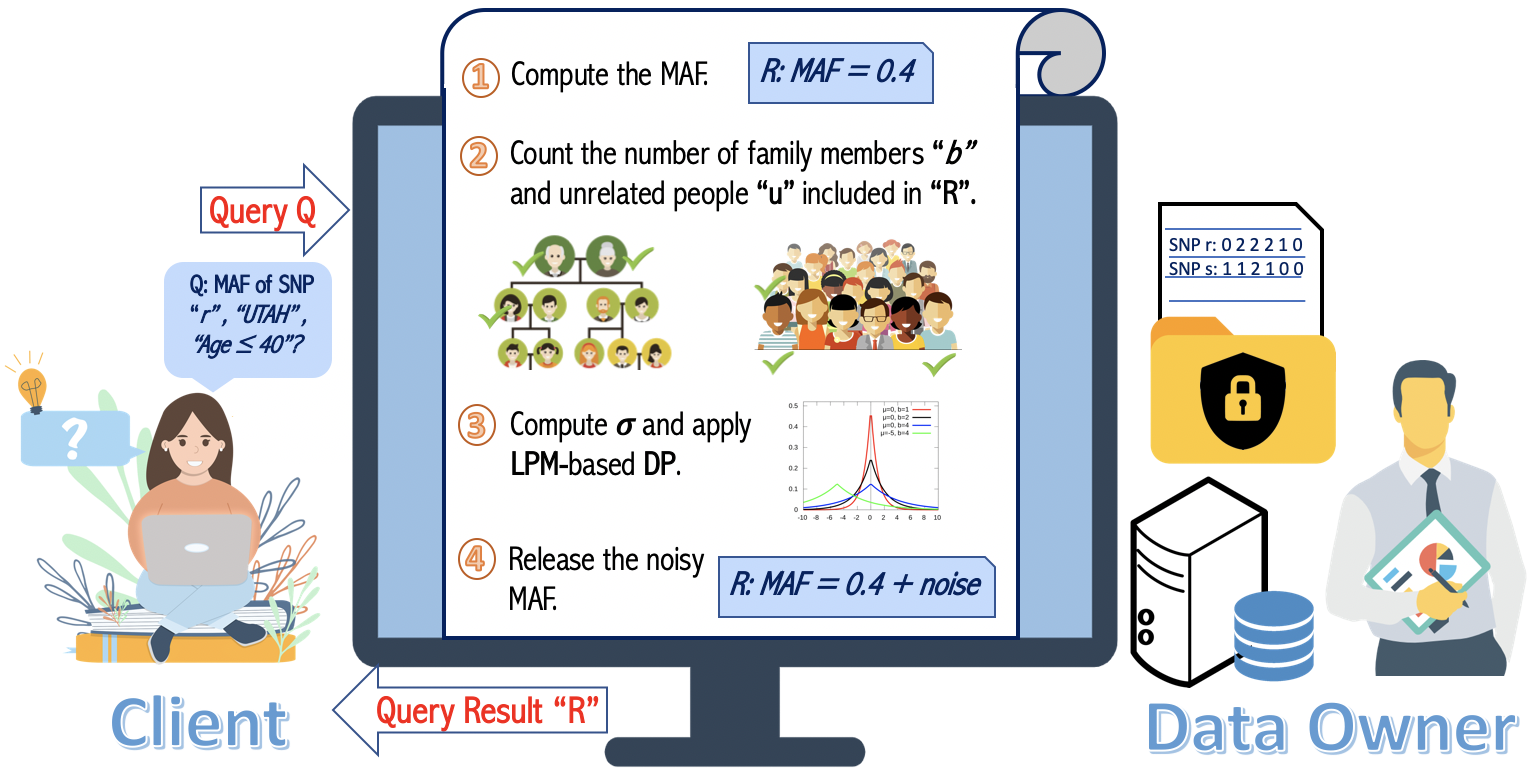}
\caption{Our proposed GenShare model} 
\label{fig1}
\end{figure*}
\section{Proposed Method} \label{MethodSec}
As discussed in Section~\ref{RelatedWorkSec}, some researchers have proposed general mechanisms to tackle the degradation in the privacy guarantees of DP that happens on account of the dependency between database tuples~\cite{kifer2012rigorous}. However, this privacy risk has not yet been studied for statistical genomic datasets (which potentially include many dependent tuples due to dependency/correlations between genomes of individuals that have family ties) and existing mitigation~\cite{chen2014correlated,zhao2017dependent,chen2017pegasus,liu2016dependence,almadhoun2020differential} fail to theoretically capture the statistical relationships between dependent tuples in genomic datasets, and hence resulting in sub-optimal solutions considering privacy and utility.

As a first step towards mitigation of this risk, following a similar analysis as in~\cite{liu2016dependence} (but modeling the correlations differently, i.e., joint correlations considered), we propose GenShare as a formalization of $\epsilon$-DP notion for genomic datasets with dependent tuples. Among all family trees in a dataset $D$, we denote the one with the strongest relationships (i.e., the one with the largest
aggregate kinship coefficient between any individual and the other family members) as the strongest dependent
tuple set and represent it as $B$ ($|B|= b$). We let $D$ and $D'$ be neighboring datasets with $b$ dependent tuples (i.e., among $b$ dependent tuples, $D$ and $D'$ differ in one record) if the change of one tuple value in $D$ causes change of at most $(b-1)$ tuple values in $D'$. Thus, we define GenShare for genomic datasets with dependent tuples using this notion of neighboring datasets, and to achieve the guarantees of $\epsilon$-DP, we re-formulate LPM by introducing a new fine-grained “sensitivity” definition $\varsigma$ for genomic datasets that include dependent tuples, as follows:

\begin{theorem} \label{DPtheorem}
For a dataset $\textit{D}$ with \textit{b} genomic dependent tuples, a randomized algorithm $A$ provides $\epsilon$-differential privacy for a query \textit{Q} with global sensitivity $\varsigma$, if $\textit A(D) = Q(D)+LPM(\varsigma / \epsilon)$.
\end{theorem}

\begin{lemma}
Let \textit{$\Delta Q$} represent the global sensitivity of a query \textit{$Q$}. %(i.e., maximum difference in query results obtained from any two neighboring datasets that include uncorrelated or independent tuples). We define 
The dependent sensitivity for sharing the results of query \textit{$Q$} over a genomic dataset with dependent tuples $\varsigma  = \Delta Q + \sigma(B)$.
\end{lemma}
\begin{prooF}
To prove Theorem \ref{DPtheorem} and compute $\sigma(B)$, we consider a simple query function to publish a sanitized version $\tilde{D}$ of a dataset $D$ with $b$ dependent tuples. Among these $b$ dependent tuples, we have the participant $j$ and participants in set $\psi$, where $\psi$ may contain more than one tuple. To satisfy $\epsilon$-DP under this scenario we have:
\begin{equation} \label{e-dp}
\mathop{max}_{h, h'} {\frac{P(A([x_j^i = h, x_{\psi}^i]) = [\tilde{h}, \tilde{x_{\psi}^i}])}{P(A([x_j^i = h', x_{\psi}^i]) = [\tilde{h}, \tilde{x_{\psi}^i}])}}\leq {exp (\epsilon)}
\end{equation}

where $A$ is a randomized algorithm, $i'$ represents the sanitized version of a data point (SNP) $i$, $x_{j}^i$ represents the SNP $i$ value of individual $j$, and $x_{\psi}^i$ represents the set of SNP values of the individuals in set $\psi$, where $\psi$= $B/\ j$. Also, $h$ and $h'$ values are selected to obtain the maximum difference in the $x_j^i$ value (i.e., $h'$ = 2 if $h$ = 0 and if $h'$ = 0, $h$ = 2). This is to consider the effect of maximum change in the SNP $i$ value of $j$ participant on the values of dependent individuals in $\psi$. 

To achieve $\epsilon$-DP, we add Laplace noise proportional to the query's global sensitivity, by using a proper Laplace scale $\omega$ for the Laplace distribution, where $\omega$ = $\Delta Q / \epsilon$. Our goal is to find a proper scaling factor $\omega$ when sharing statistics from a dataset with dependent tuples by changing the original global sensitivity $\Delta Q$ to $\varsigma$.
By transforming the left-hand side of Equation \ref{e-dp} using the law of total probabilities, we have:
\begin{multline} \label{e1-dp}
\mathop{max}_{h, h'}{\frac{P(A([x_j^i = h, x_{\psi}^i]) = [\tilde{h}, \tilde{x_{\psi}^i}])}{P(A([x_j^i = h', x_{\psi}^i]) = [\tilde{h}, \tilde{x_{\psi}^i}])}}\\
\leq \mathop{max}_{h, h'} {\frac{P(\tilde{x_j^i} = \tilde{h}| x_j^i = h)}{P(\tilde{x_j^i} = \tilde{h}| x_j^i = h')}}\\
\cdot \mathop{max}_{h, h'} {\frac{\sum_{\bar{a} \in \mathbb{A}} P(x_{\psi}^i = \bar{a}|  x_j^i = h)P(\tilde{x_{\psi}^i} = \tilde{\bar{a}}| x_{\psi}^i = \bar{a})}{\sum_{\bar{a} \in \mathbb{A}} P(x_{\psi}^i = \bar{a} |  x_j^i = h')P(\tilde{x_{\psi}^i} = \tilde{\bar{a}}| x_{\psi}^i = \bar{a})}}\\
\leq \mathop{max}_{h, h'} {\frac{P(\tilde{x_j^i} = \tilde{h}| x_j^i = h)}{P(\tilde{x_j^i} = \tilde{h}| x_j^i = h')}} \cdot {\mathop{max}_{h, h'} {\frac{\sum_{\bar{a} \in \mathbb{A}} f(\bar{a}) \cdot P(x_{\psi}^i = \bar{a}|  x_j^i = h)}{\sum_{\bar{a} \in \mathbb{A}} f(\bar{a}) \cdot P(x_{\psi}^i = \bar{a} |  x_j^i = h')}}}  
\end{multline}
\vspace{20pt}

Here, $\bar{a}$ is a vector representing the values of the SNPs in $x_{\psi}^i$. $\mathbb{A}$ includes the set of vectors for potential values of $\bar{a}$ (considering Mendel's law and the relationships of the dependent tuples in the dataset). Also, $f(\bar{a})$ is a function that computes the sum of SNP values in $\bar{a}$. To compute the potential values in $\bar{a}$, we develop probabilistic models representing the evolution of an SNP value over multiple generations. For this, based on Mendel’s law%and minor allele frequencies
, we find the family relationships between individuals and compute the probabilities of moving from one SNP value to another, from one generation to the next.%Mendelian inheritance probabilities for a SNP given all the SNP genotypes of the parents are explained in Section in the Supplementary Materials.
The right-hand side of Equation \ref{e1-dp} contains two terms: the first left term considers the change in the SNP $\textit{i}$ of individual $\textit{j}$ from the value $\textit{h}$ to $\textit{h'}$, and the second right term that considers the change in the SNP $\textit{i}$ of individuals in $\textit{$\psi$}$ (due to the dependency between $\textit{j}$ and individuals in $\textit{$\psi$}$) given the change in $x_j^i$ from the value $\textit{h}$ to $\textit{h'}$. For the first left term of the right-hand side of Equation \ref{e1-dp}, we have:

\begin{multline} \label{e2-dp}
{\mathop{max}_{h, h'} {\frac{P(\tilde{x_j^i} = \tilde{h}| x_j^i = h)}{P(\tilde{x_j^i} = \tilde{h}| x_j^i = h')}} }
= \mathop{max}_{h, h'} {\frac{exp (\frac{\left \| \tilde{h}- h \right \|}{\omega})}{exp (-\frac{\left \| \tilde{h}- h' \right \|}{\omega})}}\\
\leq \mathop{max}_{h, h'} exp (\frac{\left \| h - h' \right \|}{\omega})
\leq {exp (\frac{\Delta x_j^i }{\omega})}
\end{multline}

 where $\Delta x_j^i$ represents $\Delta Q$ which is the maximum change in $x_j^i$ from the value $\textit{h}$ to $\textit{h'}$. If we ignore the second right term of the right-hand side of Equation \ref{e1-dp}, and combine the remaining of Equation \ref{e-dp} and Equation \ref{e1-dp} , then we have:

 \begin{equation}\label{e6-dp}
 {exp (\frac{\Delta x_j^i }{\omega})} = {exp (\epsilon)}
 \end{equation}
 
The scale $\omega$ for the Laplace distribution is:
 $\omega = \frac{\Delta x_j^i }{\epsilon}$
which is compatible with the Laplace scale in the standard DP mechanism. To study the effect of the the maximum change in an individual $j$'s data on \textit{b-1} dependent tuples (in $\psi$), we focus on the second right term of the right-hand side of Equation \ref{e1-dp} to define \textit{$\sigma$} as follows:

\begin{equation} \label{e3-dp}
exp(\frac{\sigma}{\omega}) =
{\mathop{max}_{h, h'} {\frac{\sum_{\bar{a} \in \mathbb{A}} f(\bar{a}) \cdot P(x_{\psi}^i = \bar{a}|  x_j^i = h)}{\sum_{\bar{a} \in \mathbb{A}} f(\bar{a}) \cdot P(x_{\psi}^i = \bar{a} |  x_j^i = h')}}}
\end{equation}

Combining Equation \ref{e-dp}-\ref{e3-dp}, we have:
\begin{multline}\label{e4-dp}
%\frac{P(A([x_j^i = h, x_{\psi}^i]) = [\tilde{h}, \tilde{a}])}{P(A([x_j^i = h', x_{\psi}^i]) = [\tilde{h}, \tilde{a}])}
\mathop{max}_{h, h'} {\frac{P(A([x_j^i = h, x_{\psi}^i]) = [\tilde{h}, \tilde{x_{\psi}^i}])}{P(A([x_j^i = h', x_{\psi}^i]) = [\tilde{h}, \tilde{x_{\psi}^i}])}}
\\
\leq exp (\frac{\Delta x_j^i }{\omega}) \cdot exp(\frac{\sigma}{\omega})
= exp (\frac{(\Delta x_j^i + \sigma)}{\omega})
\end{multline}

Therefore, we represent the dependent sensitivity for sharing the results of query \textit{$Q$} over a genomic dataset with dependent tuples as 
$\varsigma$  = $\Delta$ $x_j^i$ + $\sigma(B)$ = $\Delta Q + \sigma (B)$.
\end{prooF}

We derive the dependent sensitivity $\sigma(B)$ as:
\begin{equation}
 \sigma (B) =    \max_{j\in\mathbf{B}} log{
 %\frac{\sum_{\psi \in \mathbf{B} \setminus j} \sum_{k\in\{0,1,2\}} k \cdot P(x_{\psi}^i = k|  x_j^i = h)}{\sum_{\psi \in \mathbf{B} \setminus j} \sum_{k\in\{0,1,2\}} k \cdot P(x_{\psi}^i = k |  x_j^i = h')}
 \frac{\sum_{\bar{a} \in \mathbb{A}} f(\bar{a}) \cdot P(x_{\psi}^i = \bar{a}|  x_j^i = h)}{\sum_{\bar{a} \in \mathbb{A}} f(\bar{a}) \cdot P(x_{\psi}^i = \bar{a} |  x_j^i = h')}
 } 
 \cdot \omega
\end{equation} 

In practice, depending on over which individuals a query is computed, first the strongest dependent tuple set $B$ among such individuals is determined, and then, the corresponding dependent sensitivity $\sigma (B)$ is computed. Furthermore, we observe that the inference power of an adversary may be affected by the number of dependent tuples (i.e., family members) and independent tuples (i.e., unrelated members) included in the query results. Hence, in our sensitivity analysis, $\omega$ (i.e., LPM scale) value can be neatly chosen to find the adequate value of $\sigma (B)$. We show our heuristic analysis on how to choose $\omega$ in Section~\ref{ExperimentalResultsSec}.

\subsection{Use Case}
To clarify our previous computations, here we consider a simple query function to publish a sanitized version $\tilde{D}$ of a dataset $D$ with $b$ dependent tuples. Among these $b$ dependent tuples we have the participants j and k, and o, where k and o $\in \psi$. To satisfy $\epsilon$-DP for genomic datasets with dependent tuples we have:
\begin{equation} \label{0e-dp}
{\mathop{max}_{h, h'} {\frac{P(A([x_j^i = h, x_k^i, x_o^i]) = [\tilde{h}, \tilde{a}, \tilde{b}])}{P(A([x_j^i = h', x_k^i, x_o^i]) = [\tilde{h}, \tilde{a}, \tilde{b}])}}} \leq {exp (\epsilon)}
\end{equation}

By transforming the left-hand side of Equation \ref{0e-dp} using the law of total probabilities, we have:
\begin{multline} \label{0e1-dp}
\mathop{max}_{h, h'} {\frac{P(A([x_j^i = h, x_k^i, x_o^i]) = [\tilde{h}, \tilde{a}, \tilde{b}])}{P(A([x_j^i = h', x_k^i, x_o^i]) = [\tilde{h}, \tilde{a}, \tilde{b}])}}\\
\leq {\mathop{max}_{h, h'} {\frac{P(\tilde{x_j^i} = \tilde{h}| x_j^i = h)}{P(\tilde{x_j^i} = \tilde{h}| x_j^i = h')}}}\\
\cdot{\mathop{max}_{h, h'} {\frac{\sum_{a \in\{0,1,2\}} \sum_{b \in\{0,1,2\}} P(x_k^i = a, x_o^i = b|  x_j^i = h)}{\sum_{a \in\{0,1,2\}} \sum_{b \in\{0,1,2\}} P(x_k^i = a, x_o^i = b|  x_j^i = h')}}}
\cdot{{\frac{P(\iota)}{P(\iota)}}}\\
(\iota = \tilde{x_k^i} = \tilde{a}, \tilde{x_o^i} = \tilde{b}| x_k^i = a, x_o^i = b) \\
\leq {\mathop{max}_{h, h'} {\frac{P(\tilde{x_j^i} = \tilde{h}| x_j^i = h)}{P(\tilde{x_j^i} = \tilde{h}| x_j^i = h')}}}\\
\cdot {\mathop{max}_{h, h'} {\frac{\sum_{a \in\{0,1,2\}} \sum_{b \in\{0,1,2\}} P(x_k^i = a, x_o^i = b|  x_j^i = h) \cdot (a+b) }{\sum_{a \in\{0,1,2\}} \sum_{b \in\{0,1,2\}} P(x_k^i = a, x_o^i = b|  x_j^i = h') \cdot (a+b)}}}
\end{multline}
\vspace{10pt}

Therefore, we derive the dependent sensitivity $\sigma$ as:

\scalebox{0.90}{
 $\sigma =    \max_{j\in\mathbf{B}} log{
 \frac{\sum_{a \in\{0,1,2\}} \sum_{b \in\{0,1,2\}} P(x_k^i = a, x_o^i = b|  x_j^i = h) \cdot (a+b) }{\sum_{a \in\{0,1,2\}} \sum_{b \in\{0,1,2\}} P(x_k^i = a, x_o^i = b|  x_j^i = h') \cdot (a+b)}} \cdot \omega$} 
 %\end{equation} 
 \begin{figure*}
 \centering
\includegraphics[width=0.8\textwidth]{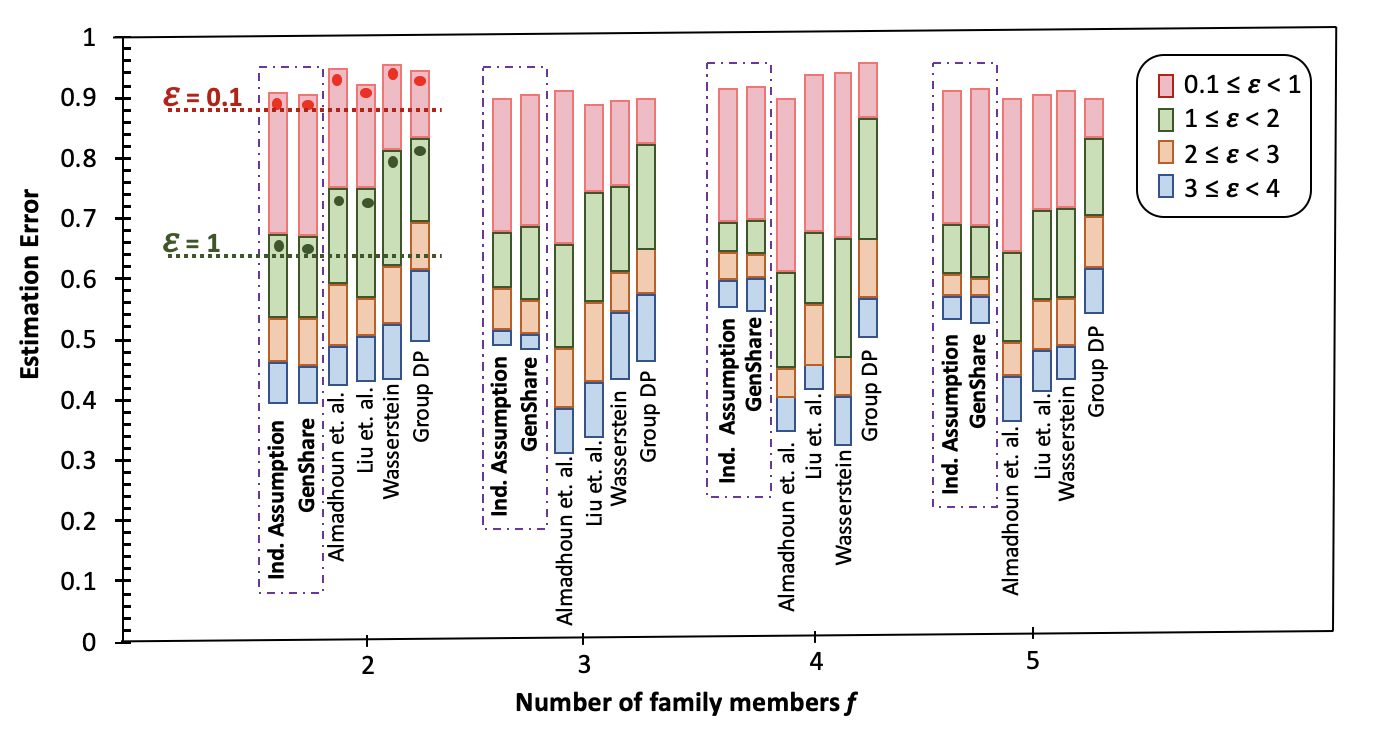}
\caption{The effect of including only the target and his relatives from MC family in the count query results, on the adversary's estimation error of inferring the target's SNPs values. Using a range of $\epsilon$ values, we compare our model “GenShare" with 5 existing mechanisms. We provide the data points of the estimation error when we use $\epsilon$ = 0.1 and 1 for a query with 2 family members.} \label{fig2}
\end{figure*} 
\subsection{GenShare Model}
Let dataset $\textit{D}$ includes $n$ individuals and $\textit{m}$ SNPs. We assume a statistical query (e.g. MAF) is computed over $\textit{q}$ participants in $D$, including a target $j$ and other $p$ dataset participants ($\textit{q = 1+p}$). Set $\mathbf{F}$ ($\mathbf{|F|}= f \leq d$) includes individuals from the same family (i.e., target $j$ and his/her family members), and set $\mathbf{U}$ ($\mathbf{|U|}= u$) includes the other unrelated members (non-relatives) in the dataset. We show the overview of our proposed GenShare model in Figure~\ref{fig1}. 
The entity which collects/generates the genomic dataset is the “data owner” and the data owner can share statistics about its dataset with a client (i.e., researcher or physician). This is a common way to share research findings. Following the attack scenario proposed by~\cite{almadhoun2020differential}, to limit the number of dataset members included in the query result, the client (or adversary) sends its query specified by some demographic properties (e.g., age, address). As an example, we consider here the MAF query by the client (or adversary).  First, the data owner computes the result of the query on the dataset, and meanwhile, he determines the number of family members $f$ and unrelated members $u$ included in the query results. Based on that, the data owner computes $\sigma (B)$ and then applies LPM to the query results, then he sends them to the client.
Data owner reports (i) the query result (MAF of all SNP values for the dataset participants that are considered in the query computation) and (ii) the number of dataset participants that are used to compute the query results ($\textit{q}$).

%% file: 4-Results.tex
\section{Settings and Evaluation} \label{EvaluationSec}
To evaluate the privacy performance of our proposed model GenShare, we {use} the correctness metric over a real-world statistical genomic dataset to show the robustness of GenShare. We next discuss our evaluation in detail.
\subsection{Dataset Discription} \label{DatasetDescriptionSec}
We combine three statistical genomic datasets that include genomic data of 1) family members and, 2) unrelated members (non-relatives). Our final genomic datasets contain the partial DNA sequences from:
\begin{itemize}
   \item CEPH/Utah Pedigree 1463~\cite{drmanac2010human}: to obtain the genotypes of 10 family members (originally 17 members) from variant call format (VCF) files. 
   \item Manuel Corpas (MC) Family Pedigree~\cite{corpas2013crowdsourcing}: to obtain the genotypes of a scientist named Manuel Corpas (the target in our experiments) and his 4 family members.
   \item 1000Genome phase 3 data~\cite{10002015global}: to obtain data for the unrelated individuals from the same or different population of the target and his family members. We extracted the genotypes from chromosomes 1 and 22 for 2504 participants from 23 populations using the Beagle genetic analysis package~\cite{browning2018one} (to extract the number of minor alleles for each SNP).
\end{itemize}
\subsection{Differentially private data release} \label{DPAlgorithmsSec}
In a statistical genomic dataset (e.g., GWAS) with $n$ individuals and $m$ SNPs, \cite{uhlerop2013privacy} computes the sensitivity for privacy-preserving release of cell counts as 2 (i.e., Laplace noise with scale 2/$\epsilon$), while the MAF sensitivity can be computed as $\frac{2m}{n}$ and ${\chi}^2$ statistics as $4n\over(n+2)$. \cite{johnson2013privacy} claim that adding Laplace noise with scale $2 \over \epsilon$ to the cell count of genomic dataset results in accurate ${\chi}^2$ statistics or $p$-values. In GenShare, we use these algorithms to calculate the global sensitivity of the queries $\Delta Q $.
\begin{figure}
\center
\includegraphics[width=0.45\textwidth]{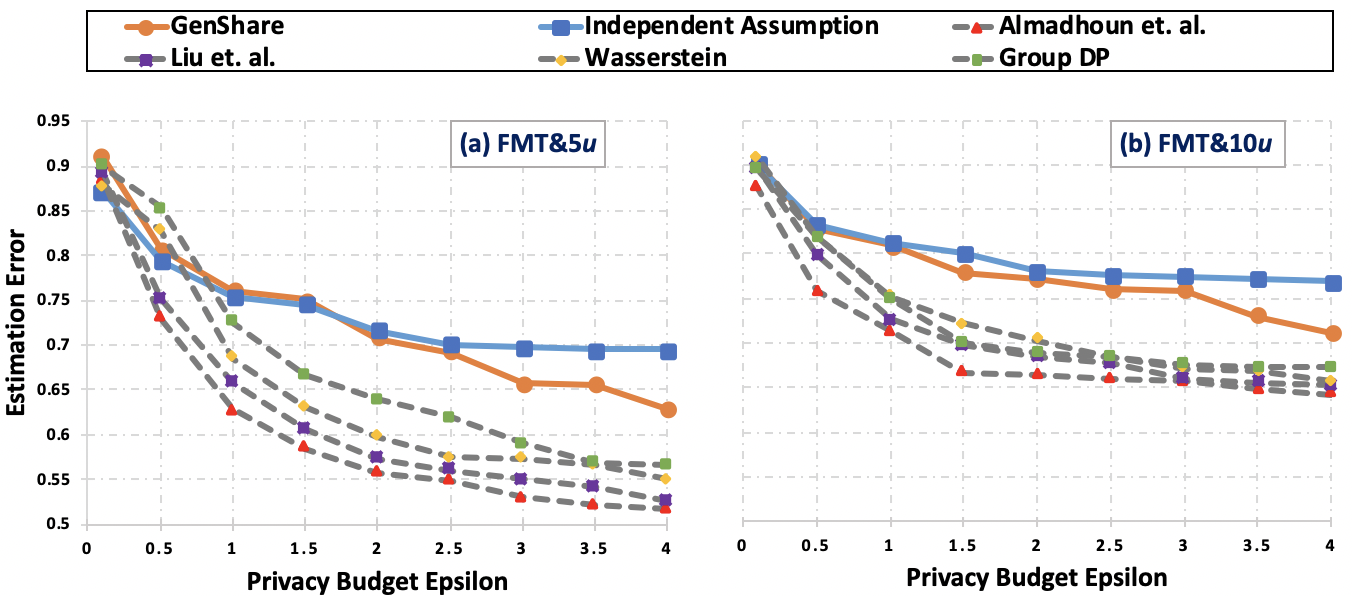}
\caption{The effect of including the target, his father and mother from MC family, and (a) 5 unrelated members (FMT5u) or (b) 10 unrelated members (FMT10u) in the count query results, on the adversary's estimation error of inferring the target's SNPs values.} \label{fig3}
\vspace{-10pt}
\end{figure}
\begin{figure*}[b]
\center
\includegraphics[width=0.9\textwidth]{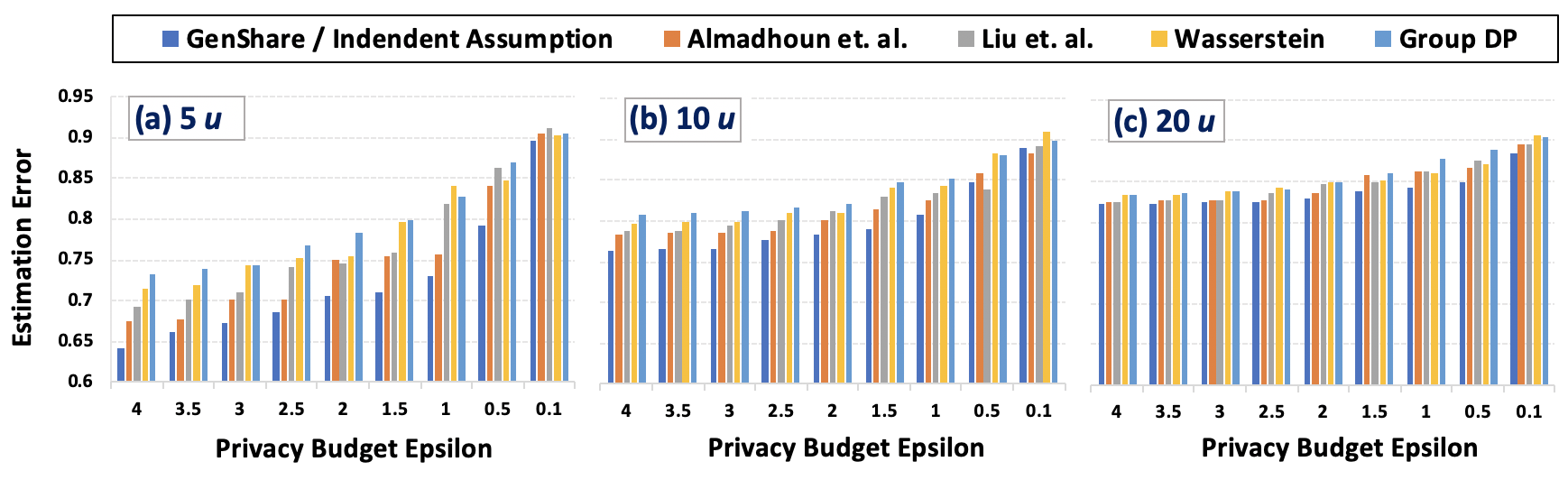}
\caption{The effect of including the target $j$, and only (a) 5 unrelated members (5u), (b) 10 unrelated members (10u), (c) 20 unrelated members (20u) in the MAF query results, on the adversary's estimation error of inferring the target's SNPs values.} \label{fig4}
\end{figure*}
\begin{figure*}[t]
\centering
\includegraphics[width=0.9\textwidth]{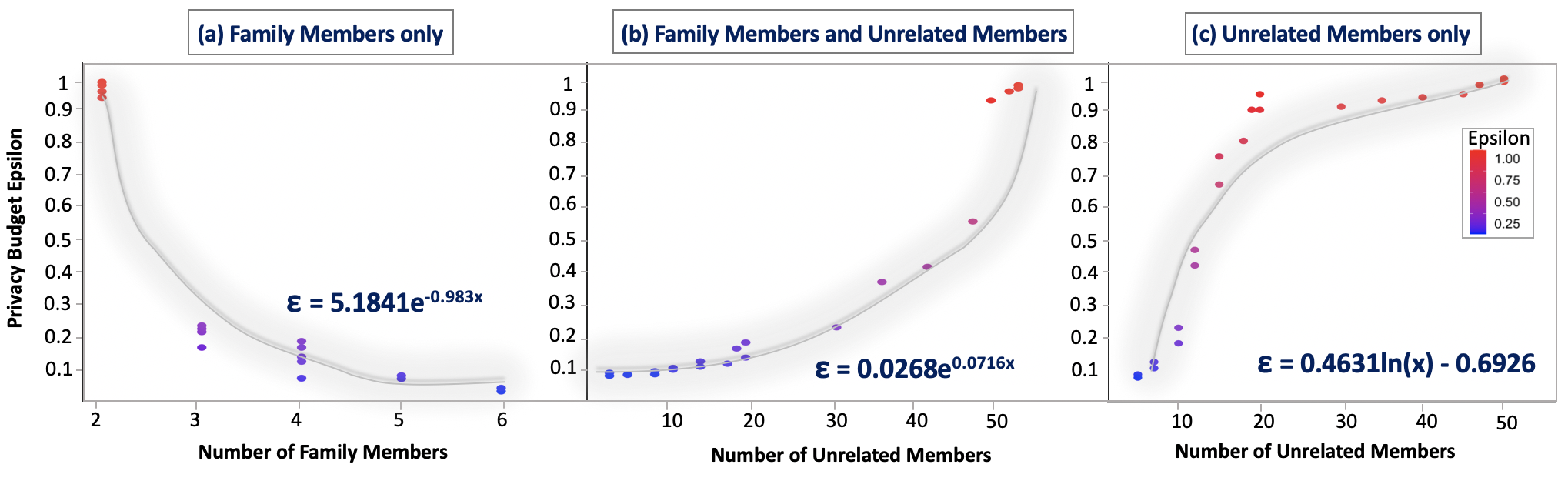}
\caption{$\epsilon$ calculations for computing $\omega$ in the sensitivity $\sigma (B)$ when the query results contain (a)family members only, (b)family members and other unrelated members, (c)unrelated members only} \label{fig5}
\end{figure*}
\subsection{Evaluation Metrics} \label{EvaluationMetricsSec}
%-\rephrase{Utility (loss) function: A function that quantifies how desirable (undesirable) a certain outcome of the system is to the user. In this work, this function takes the true query result and a perturbed query result (from a DP mechanism) as input and outputs a real number}
For evaluating GenShare, we use~\emph{correctness} metric to quantify the privacy-preserving guarantees of GenShare. \emph{Estimation error} is used to quantify the correctness by measuring the distance $\textit{Dist}$ between the true value of the SNP and the inferred value by the client (e.g., adversary). For a statistical genomic dataset $D$ with $m$ SNPs, we measure the expected estimation error $E$ as follows:
\begin{equation} \label{equ:error}
 E =\sum_{i=1}^{m}P\left ( x_j^i \mid D_{j}^{i} \right )\left | Dist\left ( x_j^i, {x}_{j}^{i} \right ) \right |
\end{equation}
Here, $x_j^i$ is the true value of SNP $\textit{i}$ for the target individual $\textit{j}$, while ${x}_{j}^{i}$ is the estimated value. We can compute the  probabilities for $x_j^i$ using the Mendelian inheritance probabilities for a SNP $i$ given all the potential SNP values (i.e., 0, 1, or 2) for $x_j^i$ (represented as $D_j^i$).
As discussed in Section~\ref{DatasetDescriptionSec}, we use a dataset $D$ to evaluate GenShare and compare it with the state-of-the-art mechanisms. $\textit{D}$ includes $\textit{n}$ individuals ($\textit{n}$= 2520) and $\textit{m}$ SNPs for each individual ($\textit{m}$ = 1000). To infer the values of these $\textit{m}$ SNPs, we repeat our experiments 10$\times$ considering 100 SNPs (i.e., 100 queries are performed) each time. 
\subsection{Experimental Results} \label{ExperimentalResultsSec}
In our evaluation, we assume that the query can include the target (e.g., individual $\textit{$j$}$) with  1) a direct family member, 2) multiple family members, or 3) multiple family members, and other unrelated individuals. We compare our model (in terms of privacy) with the existing similar work (discussed in Section~\ref{RelatedWorkSec}) such as~\cite{liu2016dependence,almadhoun2020differential,song2017pufferfish,dwork2014algorithmic}. Since Hidden Markov would not work to model kinship relations in a genomic dataset, we are not comparing our model with the mechanisms proposing Hidden Markov-based models. 
In the following, we compare our proposed model (referred to as “GenShare" in the figures) with: (i) independent assumptions (referred to as “Independent Assumption" in the figures) to show that GenShare can be proven by preventing any client from utilizing the dependencies among the dataset tuples to infer more sensitive attributes about dataset participants (in other words, we are aiming at achieving the privacy guarantees of the standard DP assuming all the participants of the dataset are independent), (ii) the proposed mitigation algorithm in~\cite{almadhoun2020differential} (referred to as “Almadhoun et. al.” in the figures), (iii) dependent sensitivity mechanism proposed in~\cite{liu2016dependence} (referred to as “Liu et. al." in the figures), (iv) Wasserstein algorithm proposed in~\cite{song2017pufferfish} (referred to as “Wasserstein" in the figures), and (v) Group DP proposed in ~\cite{dwork2014algorithmic} (referred to as “Group DP" in the figures).
%we observed that the differentially private statistics calculated based on GenShare more accurately matched the ground truth as opposed to other baseline mechanisms
%our mechanismalso achieves smaller error probabilities compared to the exist-ing DP techniques 
%Our approach achieves the best accuracy overall, comparable to the Laplace mechanism yet considerably more accurate than the exponential mechanism.

In Figure~\ref{fig2}, we evaluate the effect of different values of the privacy budget, $\epsilon$, on the adversary's correctness in inferring the targeted $m$ SNPs considering a different number of family members included in the query results. We evaluate the estimation error using 18 different $\epsilon$ values (i.e., $\epsilon$ is not continuous, $0.1 \leq \epsilon \leq 4$) divided into 4 intervals as shown in the legend of Figure~\ref{fig2}.
%We also analyze the robustness of our proposed mechanism to the inference attack and compare it with the most similar existing work~\citep{almadhoun2019differential}.

Here, the count query (used in cohort discovery) results include the statistics from the family members only. First, we start including 1 first-degree family member (e.g., mother or father) from MC family with the target $j$. Then, we include both mother and father with the target $j$ to the query results. Third, we include father, mother, and sister in the query results. Last, we consider a second-degree family member (aunt of the target $j$) in the query results along with the father, mother, and sister of the target as shown in the (x-axis) of Figure~\ref{fig2}. We make the following key observations: (i) GenShare achieves the best privacy overall, it provides almost the same privacy guarantees (in terms of estimation error), as the query that is computed over independent tuples (i.e., independent assumption). Hence, our model succeeds in near-achieving the standard differential privacy guarantees without any degradation in terms of privacy or utility across several $\epsilon$ values. (ii) Existing techniques generally cannot optimize their schemes to achieve the required privacy and utility guarantees. They either add too much noise (e.g., f= 2 members in the figure) or degrade rigorous guarantees of privacy (e.g., as when f $\geq$ 3 members). (iii) As expected by DP, decreasing the privacy budget values (starting from $\epsilon$= 4 descending until $\epsilon$= 0.1) leads to increasing the privacy guarantees while decreasing the utility guarantees.

Next, in Figure~\ref{fig3}, we include family members (father and mother) and other unrelated members ($u$= 5 in Figure~\ref{fig3}(a) and $u$= 10 in Figure~\ref{fig3}(b)) with the target $j$ to evaluate the effect of different values of the privacy budget, $\epsilon$, on the adversary's correctness in inferring the targeted $m$ SNPs. Considering a count query, we observe that GenShare achieves better privacy for various privacy budgets, compared to the existing techniques even when the query results include unrelated members, as illustrated in Figure~\ref{fig3}.

Figure~\ref{fig4} shows that GenShare is equivalent to DP mechanism when the query results only include unrelated members, unlike the existing techniques~\cite{liu2016dependence,song2017pufferfish,dwork2014algorithmic}, which compute the dependent sensitivity based on the number of dependent tuples in the dataset, ignoring whether these dependent tuples are included in the query or not. 

In our sensitivity analysis in Section~\ref{MethodSec} we observe that the inference power of an attacker decreases with an increasing number of independent tuples in the query computation. Hence, $\omega$ (i.e., LPM scale = $\Delta Q / \epsilon$) value can be neatly chosen to find the adequate value of $\sigma (B)$ considering the number of dependent and independent tuples in the query computation. Since the $\Delta Q$ (i.e., the query sensitivity) is computed considering the query type (illustrated in Section~\ref{DPAlgorithmsSec}), the data owner in our model can compute the $\epsilon$ value in $\omega$ based on the number of family members or unrelated members included in the query result, as shown in Figure~\ref{fig5}. As expected, adding more unrelated members to the query results leads to more precise sensitivity computations until reaching the sensitivity of the standard DP mechanism (i.e., $\varsigma$ = $\Delta Q$).

\begin{figure}
\vspace{-20pt}
\center
\includegraphics[width=0.4\textwidth]{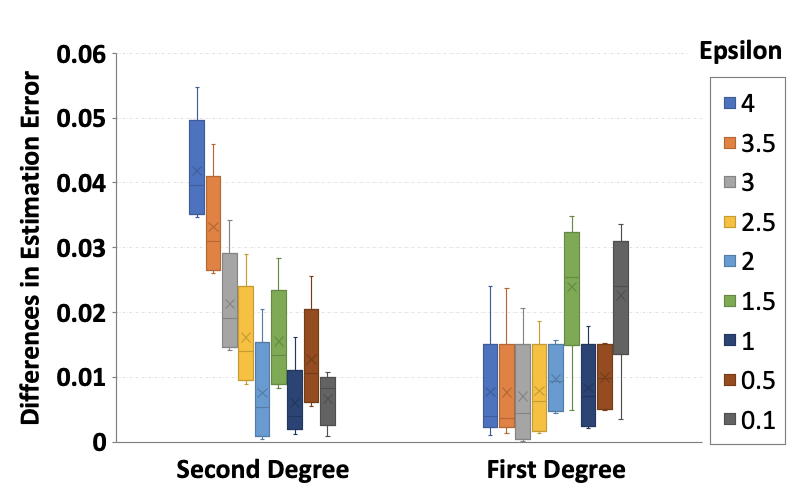}
\caption{The differences from “Independent Assumption" privacy guarantees (in terms of estimation error), considering range of $\epsilon$ values and different cases of including first-degree and second-degree relatives in the query computations.} \label{fig6}
\vspace{-20pt}
\end{figure}

\begin{figure}
\center
\includegraphics[width=0.4\textwidth]{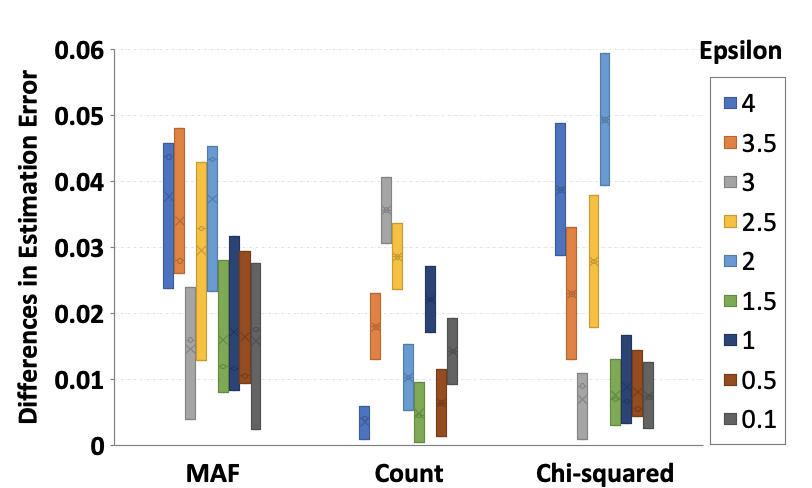}
\caption{Comparison between applying GenShare for count, MAF, and ${\chi}^2$ queries. GenShare reduces the differences from “Independent Assumption" privacy guarantees (in terms of estimation error), considering different $\epsilon$ values and the 3 query types} \label{fig7}
\end{figure}

Next, we compare the performance of GenShare when first-degree or second-degree family members (from MC and UTAH families) are included in the query computations with the target. Our results show the robustness of GenShare regardless of the degree of familial relationship between the dataset tuples. The differences in privacy guarantees between GenShare and the “Independent Assumption" do not exceed 5\% across a range of privacy parameters $\epsilon$, with respect to estimation error (Figure~\ref{fig6}). 
%In accordance with the results of~\cite{almadhoun2020inference}, the most accurate inference of the adversary is achieved when the query computation includes target $j$ along with his father and mother (Figure~\ref{fig:02}(c)). Including a second-degree family member as in (Figure~\ref{fig:02}(d)) can enlarge the range of possible SNP values for the target, and hence make it more difficult to accurately infer the correct SNP value with a high probability.
Finally, we compare the performance of GenShare for different query types, e.g., count, MAF, and ${\chi}^2$ tests. As expected, we observe that the differentially private statistics calculated based on GenShare provide accurate and near-optimal matching to the privacy guarantees of DP with “Independent  Assumption", with a difference up to 6\% in terms of estimation error across a range of privacy parameters $\epsilon$ (Figure~\ref{fig7}).
Overall, our results illustrate the theoretical boundaries of leveraging LPM-DP for mitigating the “tuples dependency" privacy risk in genomic query-answering systems. GenShare is vital for genomic data sharing and in a broader sense, it will also have implications for medical data sharing. Considering i) the importance of sharing statistical genomic and medical datasets (which is the aim that many institutes are seeking to achieve) for high-impact medical research (e.g., NIH recently awarded \$73 million to collect and archive the information of genes and genomic variants for precision medicine~\cite{nihaward}) and, ii) the sensitivity of the (personal) information in these datasets (especially there is a high probability to have families in these genomic datasets), data owners should be very careful when sharing data related to such datasets.
Moreover, %considering massive amounts of centralized or distributed clinical and genetic data centers, 
GenShare can be utilized to provide strong insights to several clients from different parties about each other’s datasets (e.g., before they exchange datasets for joint research). Such privacy-preserving sharing mechanism may be helpful to accelerate the data sharing process across researchers, especially with the worldwide strict regulations of data protection  for sharing and exchanging data.

%% file: 5-Conclusion.tex
\section{Conclusion} \label{ConSec}
Differential privacy provides a theoretical notion of privacy that provides formal guarantees that the distribution of query results changes
slightly with the addition or removal of a single tuple in the dataset. However, privacy guarantees of DP-based solutions are based on the
assumption that all tuples in the dataset are independent. In reality, genomic data from different individuals may be dependent according to the genomic interactions due to the familial ties between them. In this paper, we propose GenShare to provide countermeasures against privacy risks due to dependent tuples in the statistical genomic datasets. To achieve the privacy and utility guarantees theoretically provided by DP, GenShare captures the joint statistical relationships between dependent tuples in the genomic datasets. Our results show that GenShare provides a significant improvement in the privacy and utility guarantees over existing mechanisms across a range of privacy parameters $\epsilon$. All of these contributions will benefit the medical and genomics research community, in the long run, and realize the promise of privacy-preserving access to the genomic datasets that are relied upon in future health information exchange systems. There are several directions that merit further research. It may be possible for us to consider: 1) more concepts in differential privacy, such as local sensitivity, 2) complex tasks and applications such as federated machine learning, 3) different settings e.g., larger number of queries or composing multiple queries.

% To compute the potential values in $\bar{a}$, we developed probabilistic models representing the evolution of a SNP value over multiple generations. \hl{For this, we represented the family relationships between individuals as a Bayesian network and we used a Markov chain to represent the probabilities (based on Mendel’s law and minor allele frequencies) of moving from one SNP value to another from one generation to the next.}